\documentclass{IEEEtran4PSCC}

\ifCLASSINFOpdf
   \usepackage[pdftex]{graphicx}

\else

   \usepackage[dvips]{graphicx}

\fi

\usepackage[cmex10]{amsmath}
\allowdisplaybreaks[2]     
\usepackage{array}

 \usepackage[caption=false,font=footnotesize,labelfont=sf,textfont=sf]{subfig}

\usepackage{cite}
\usepackage{amsmath,amssymb,amsfonts}
\usepackage{algorithmic}
\usepackage{graphicx}
\usepackage{textcomp}
\usepackage{xcolor}
\def\BibTeX{{\rm B\kern-.05em{\sc i\kern-.025em b}\kern-.08em
    T\kern-.1667em\lower.7ex\hbox{E}\kern-.125emX}}

\usepackage{svg}
\usepackage{algorithm}
\usepackage{arydshln}
\usepackage{url}
\usepackage{multirow} 
\usepackage{booktabs} 
\usepackage{mathrsfs}
\usepackage{siunitx}
\usepackage{pdfpages}
\usepackage[colorlinks=true,linkcolor=black,anchorcolor=black,citecolor=black,urlcolor=black,breaklinks=true]{hyperref}
\usepackage[caption=false,font=footnotesize]{subfig}
\usepackage{textcomp} 

\usepackage{soul}

\makeatletter
\let\old@ps@headings\ps@headings
\let\old@ps@IEEEtitlepagestyle\ps@IEEEtitlepagestyle
\def\psccfooter#1{%
    \def\ps@headings{%
        \old@ps@headings%
        \def\@oddfoot{\strut\hfill#1\hfill\strut}%
        \def\@evenfoot{\strut\hfill#1\hfill\strut}%
    }%
    \def\ps@IEEEtitlepagestyle{%
        \old@ps@IEEEtitlepagestyle%
        \def\@oddfoot{\strut\hfill#1\hfill\strut}%
        \def\@evenfoot{\strut\hfill#1\hfill\strut}%
    }%
    \ps@headings%
}
\makeatother

\psccfooter{%
        \parbox{\textwidth}{\hrulefill \\ \small{24th Power Systems Computation Conference} \hfill \begin{minipage}{0.2\textwidth}\centering \vspace*{4pt} \includegraphics[scale=0.06]{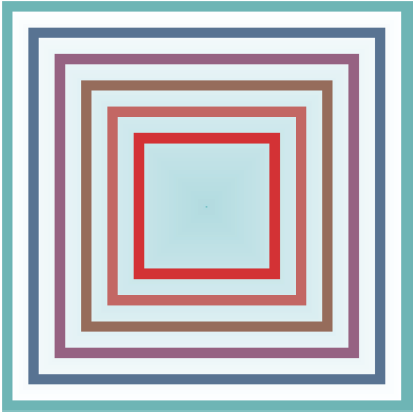}\\\small{PSCC 2026} \end{minipage} \hfill \small{Limassol, Cyprus --- June 8-12, 2026}}%
}

\begin{document}

\title{Dynamic Network Prices for Prosumer-aware Hosting Capacity Management\\}

\author{\IEEEauthorblockN{Jiawei Zhang\IEEEauthorrefmark{1},
Gregor Verbi\v{c}\IEEEauthorrefmark{1},
Frederik Geth\IEEEauthorrefmark{2},
Mohsen Aldaadi\IEEEauthorrefmark{3}, 
Rahmat Heidari\IEEEauthorrefmark{2} and
Julio Braslavsky\IEEEauthorrefmark{4}}
\IEEEauthorblockA{\IEEEauthorrefmark{1} School of Electrical and Computer Engineering\\
The University of Sydney,
Sydney, Australia\\}
\IEEEauthorblockA{\IEEEauthorrefmark{2} 
School of Electrical Engineering and Computer Science \\
The University of Queensland,
Springfield, Australia\\ }
\IEEEauthorblockA{\IEEEauthorrefmark{3} Department of Electrical Engineering\\
Islam University of Madinah,
Saudi Arabia\\}
\IEEEauthorblockA{\IEEEauthorrefmark{4} CSIRO Energy,
Newcastle, Australia\\ }
}

\maketitle

\newcommand{\varcolor}{black}

\begin{abstract}

The fast uptake of distributed energy resources (DERs) 
presents increasing challenges for managing hosting capacity in distribution networks. Existing solutions include direct load control, 
operating envelopes, 
and price-based control through dynamic \textit{energy} prices.
Despite their effectiveness, these methods often rely on assumed prosumer behavioural patterns and overlook prosumers' desire to retain control over their devices. Additionally, current fixed or Time-of-Use (ToU) prices are based on spatial and temporal averages, having limited impact on network conditions \textcolor{black}{and DER operation}. 
To address these limitations, this paper proposes a bilevel optimisation framework that explicitly models prosumer decision-making in the design of dynamic \textit{network} prices. The upper level represents the distribution system operator (DSO), setting network prices under cost-recovery and network constraints, while the lower level models prosumers optimising DER operation in response. The proposed framework preserves customer prerogative, enhances DER flexibility, and offers actionable insights for network hosting capacity management and the evolution of network tariff structures under high DER penetration.

\end{abstract}


\section{Introduction}
\label{introduction}

\subsection{Background and Motivation}
Low-voltage (LV) distribution systems were traditionally built with a `fit-and-forget' approach, designed to deliver power from substations to end users but offering limited monitoring or control capabilities\cite{Koirala_2022_RSER}.
The adoption of distributed energy resources (DERs), including residential rooftop solar photovoltaic (PV) systems, household battery energy storage systems (BESS), electric vehicle supply equipment (EVSE) and flexible loads \cite{2024_Roadmap}, result in two-way power flows and create thermal and voltage problems that strain the distribution networks \cite{Ismael_2019_RenewableEnergy}, and network upgrades are costly and slow.
As a simpler alternative, distribution utilities in Australia have enforced fixed import and export limits at customer connection points \cite{ARENA_2022_DOE}. Although effective in mitigating risks such as excessive PV exports during midday hours, this method is rigid and conservative, since the limits are determined based on infrequent worst-case scenarios. 
Consequently, there is a need for more adaptive strategies that enable efficient DER participation while leveraging available network capacity to maintain secure and reliable operation.

\subsection{\textcolor{\varcolor}{Price-based Hosting Capacity Management}}
Hosting capacity can be defined as the maximum amount of DER that can be accommodated by a network without breaching its operational performance limits. Various approaches have been proposed to manage network hosting capacity, including direct load control and dynamic operating envelopes (DOEs), which specify limits on power flows that vary with time and location, reflecting the available capacity of local distribution networks or the power system as a whole \cite{ARENA_2022_DOE}.

\textcolor{black}{Independent to DOE developments,} recent research has explored the use of price signals as a means of distribution network hosting capacity management. 
The increasing penetration of renewable generation and the emergence of price-elastic consumer behaviour provide the foundation for implementing dynamic pricing mechanisms.
For example, Ausgrid's Project Edith is piloting a dynamic pricing scheme designed to deliver fairer tariffs for customers who adjust their energy consumption in ways that support the network \cite{Edith_2022_ENEA}. The pricing model incorporates real-time conditions such as weather, solar generation, location-specific load, and time-of-day.

\textcolor{black}{
Alongside these trials, researchers have proposed a range of pricing designs. Extending the concept of locational marginal pricing from transmission-level wholesale markets, distribution locational marginal pricing (DLMP) incorporates distribution-network constraints directly into price formation. Study \cite{Papavasiliou_2018_IEEESmartGrid} compares three DLMP formulations for radial networks and finds that the marginal-losses approach aligns best with physical intuition. 
In \cite{Bai_2018_IEEEPowerSystems}, DLMPs are derived within a day-ahead distribution market-clearing model; the resulting prices are decomposed into components—marginal costs of active and reactive power, congestion, voltage support, and losses—providing location-specific signals that incentivise DERs to help manage congestion and voltage.
Similarly, \cite{Iacopo_2021_Omega} proposes a nodal pricing framework where network congestion is explicitly reflected in the price. Fixed and nodal prices coexist, enabling optimal tariffs for investment cost recovery while accommodating both flexible and non-flexible consumers within local distribution areas. 
However, applying such pricing schemes in practice can be challenging. By definition, DLMPs differ across electrical locations, raising fairness concerns as customers at different points in the network would pay different prices. Moreover, they imply that the DSO also operates the market, which is inconsistent with the Australian regulatory environment where DNSPs are separated from retailers. In addition, their practicality is limited, since implementation would require each residential customer to bid into the market.}

\textcolor{black}{Other approaches to price design and network management include the use of fuzzy logic to determine dynamic feed-in tariff which incorporates technical factors, incentivising households to export a greater share of generation during periods of feeder stress \cite{Hayat_2019_IEEEIndustrialInformatics}. In \cite{Alahmed_2025_IEEEContrlNS}, an OE-aware community market mechanism that employs two-part pricing is proposed, consisting of a threshold-based dynamic price and a fixed reward. While network constraints are captured through community-level OEs, DER curtailment is not considered, which could add more complexity to the decision-making of prosumers and operators. Finally, \cite{Afshin_2025_SEGAN} introduces a two-stage framework for managing DER and network power exchanges with the aim of maximising social welfare through coordinated power flows, dynamic consumption pricing, and feed-in tariffs. Line and voltage breaches are explicitly checked in the framework. However, privacy remains a concern, as the method requires that data be shared every hour with high-level decision makers or third parties.
}

\subsection{\textcolor{\varcolor}{Tariff design for Residential Consumers}}
\begin{figure}[!t]
    \centering
    \includegraphics[width=\columnwidth]{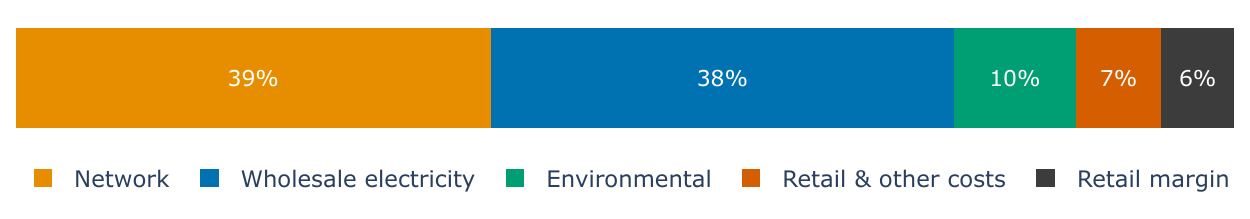}    
    \caption{Composition of an average household electricity bill in the NEM.}
    \label{fig:component_bills}
\end{figure}

\begin{figure}[!t]
    \centering
    \includegraphics[width=\columnwidth]{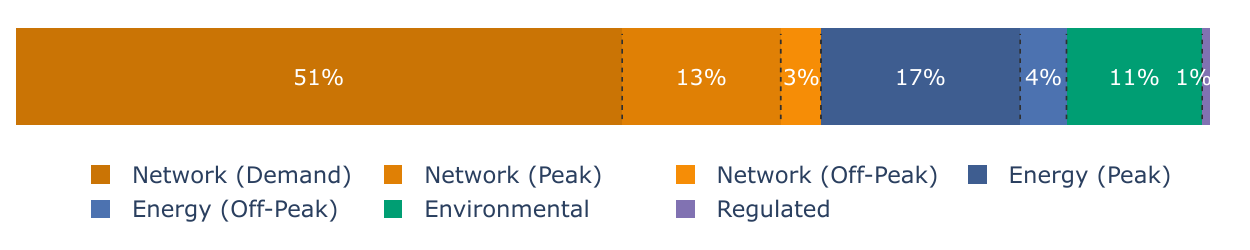}
    \caption{Composition of an example business electricity bill.}
    \label{fig:component_bills_business2}
\end{figure}

In Australia, electricity bills for residential consumers consist of wholesale energy \textcolor{black}{costs}, regulated network \textcolor{black}{charges}, environmental policy \textcolor{black}{charges}, and retailer operating expenses. Fig.~\ref{fig:component_bills} shows the share of each component in an average bill in the National Electricity Market (NEM) for 2023-2024 \cite{ACCC_2024_Tariffs}.
Network \textcolor{black}{charges}, which make up approximately 39\%, cover the operation and maintenance of network infrastructure; they are regulated by the government through the Australian Energy Regulator. A further 38\% comes from wholesale energy \textcolor{black}{costs}, which reflect the purchase of electricity from the wholesale market, which varies with market outcomes and hedging strategies. The environmental \textcolor{black}{charges} are due to renewable energy and carbon reduction schemes, contributing around 3\%. Retail \textcolor{black}{charges} capture retail operating expenses, such as billing and customer support, covering around 14\%. The remaining 6\% is the retail margin, which reflects the profit component.
For business customers, on the other hand, a typical electricity bill comprises network, energy, environmental, and regulated \textcolor{black}{charges}. An example of a realistic monthly electricity bill \cite{Acacia_bill} is shown in Fig.~\ref{fig:component_bills_business2}, where the network component comprises both demand and Time-of-Use (ToU) \textcolor{black}{charges} under the LLVT2 \textcolor{black}{tariff} from Powercor \cite{Powercor_202526}, in contrast to the fixed daily \textcolor{black}{charges} typically applied to residential customers.

The existing \textcolor{black}{retail tariff} structure for residential customers usually includes fixed, energy, and network components. The rollout of smart meters enabled the adoption of ToU \textcolor{black}{tariffs}, which distinguish between peak, shoulder, and off-peak periods to encourage customers to shift consumption to times when energy is cheaper. Network \textcolor{black}{charges} are usually applied as fixed daily components for residential customers, and distribution companies do have time-varying ones, which retailers are not obliged to pass on to residential customers. 

\begin{figure}[!t]
    \centering
    \includegraphics[width=\columnwidth]{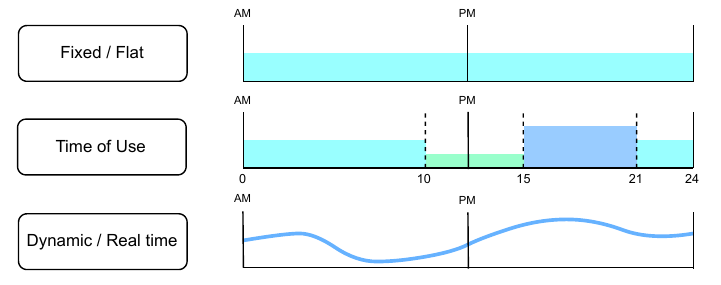}
    \caption{\textcolor{black}{Example of energy and network tariffs used in Australia}.}
    \label{fig:tariffs_plot}
\end{figure}

Fig.~\ref{fig:tariffs_plot} shows an example of \textcolor{black}{different tariff structures}. In Australia, both energy and network \textcolor{black}{tariffs} may adopt fixed or ToU structures, and dynamic energy \textcolor{black}{tariffs} also exist, such as those offered by Amber Electric whose \textcolor{black}{prices} follow wholesale market conditions. 
\textcolor{black}{
Although such tariffs might solve some distribution network problems by alleviating high demand during evening peaks, static or dynamically shaped windows  may misalign with real operating conditions. As a result, export limits, static or dynamic, are also required to ensure that network limits are not breached. 
}

\textcolor{black}{
While current ToU \textcolor{black}{tariffs} mark progress toward cost-reflective pricing, their reliance on temporal and locational averages limits their precision. Consequently, they do not incentivise targeted responses and have limited impact on local network \textcolor{black}{conditions and DER operation}.
The focus of the present work is the design of dynamic network \textcolor{black}{prices} embedding network limits to operate alongside given dynamic energy prices.
}

\subsection{\textcolor{black}{Paper Contributions and Outline}}
This paper proposes a tariff design methodology based on bi-level programming. The goal of the novel tariffs \textcolor{black}{is} to replace conventional retail tariffs, which require explicit export limits to prevent network issues, with a dynamic network price coupled with a dynamic energy price (e.g., directly linked to the wholesale market price). The dynamic price is explicitly designed to incentivise customers to shift consumption to periods when there are no network issues, as opposed to when energy is cheaper.
The proposed tariff structure may still depend on DOEs as a guardrail, but its key advantage lies in that price-based control is much more palatable to customers, as they retain their prerogative and privacy. 
The design of such network tariffs requires the bilevel model where the customer decision-making problem is explicit in the optimisation problem.
A key feature of the proposed bilevel optimisation model is that it captures the objectives of both the DSO and prosumers simultaneously, while strictly enforcing network constraints. With customers' prerogatives respected during the process, the designed tariffs encourage prosumers to operate their DER more flexibly, thereby supporting smarter management of the available network capacity.

Note that this study focuses on the redesign of network prices, with other components, including energy costs, treated as given. In addition, we assume full retail competition where customers have a contractual relationship with a retailer, but a network company is responsible for network management. Under the proposed tariff structure, the retailer passes the network price on to the customers in full to enable the mechanism to function effectively.
The contributions of the work include the following.
\begin{itemize}
    \item We propose a bilevel framework for designing dynamic network prices for hosting capacity management that separates the role of the DSO from the decisions of consumers. This formulation ensures that the objectives of both parties are respected while explicitly modelling their strategic interaction and preserving equilibrium consistency. 
    \item We explore the redesign of network prices from traditional static to dynamic ones, advancing towards more cost-reflective pricing structures and enabling targeted prosumer responses based on the local network conditions.
    \item We inform solutions for the smart management of LV network hosting capacity under high DER penetration, showing how price incentives and prosumer coordination can support reliable network operation. These economic signals effectively preserve customers' prerogatives and can be implemented with DOEs.
\end{itemize}
\textcolor{black}{The proposed framework is intended as a benchmark model for network price formation to explore the implications of price dynamics and DER operation, rather than as an online control implementation. The formulation is solved using a centralised approach, while decentralised solution techniques are beyond the scope of this work.}

The remainder of the paper is structured as follows. 
Section \ref{formulation} provides an overview of the proposed bilevel optimisation framework and describes the mathematical formulation of the problem in detail. Section \ref{casestudy} presents the case study, simulation setup and numerical results, followed by the discussion in Section \ref{discussion}. Finally, Section \ref{conclusion} concludes the paper.

\section{Mathematical Model Formulation}
\label{formulation}

This section \textcolor{black}{introduces the proposed framework and details the mathematical formulation of the bilevel optimisation. }

\subsection{Proposed Bilevel Framework}
In this work, a bilevel optimisation framework is employed to determine the dynamic network tariffs. The bilevel problem encompasses the optimisation of two nested mathematical programs, referred to as upper and lower level problems, where the upper level decision maker specifies actions subject to the optimal response of the lower-level problem. 

\begin{figure}[!t]
\centering
\includegraphics[width=0.45\textwidth]{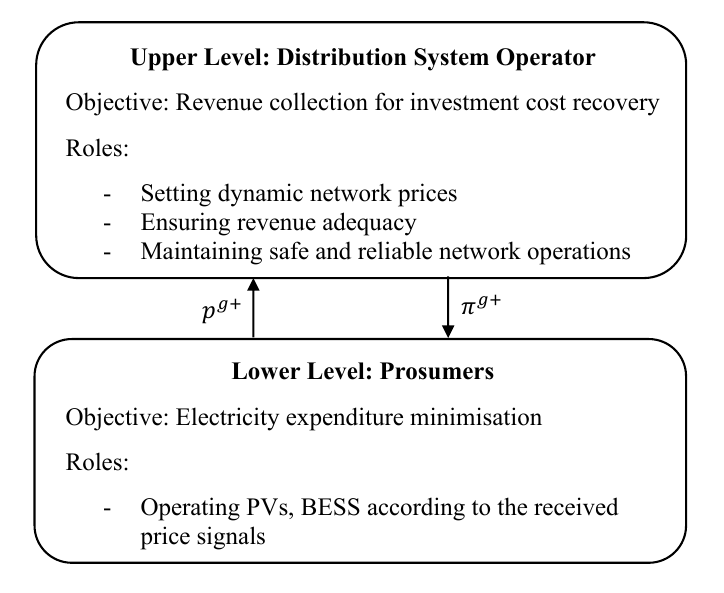}
\caption{Overview of the proposed bilevel framework.}
\label{fig:schematic}
\end{figure}
A schematic diagram of the proposed bilevel framework is shown in Fig.~\ref{fig:schematic}. \textcolor{\varcolor}{The key reason for using a bilevel framework is to account for the autonomous decision-making of prosumers, rather than assuming that the DSO can directly control behind-the-meter resources.} At the upper level, the DSO maximises its collected revenue subject to cost recovery and network constraints for power, voltage, and current. It periodically uses forecasts of prosumer demand and PV generation to set dynamic tariffs while ensuring network integrity. 
At the lower level, prosumers minimise their electricity expenditure by strategically managing their behind-the-meter DERs in response to the received price signals. 
\textcolor{black}{Note that this modelling approach assumes that the DSO has a model of customer response. In practice, there is no guarantee that customers will behave as assumed and deviations from declared operations or forecasts could occur. These uncertainties are not captured in the present formulation and will be considered in future work.}

\subsection{Optimisation Problem for DSO}
At the upper level, the DSO calculates the dynamic tariff, which is then communicated through distribution companies to the prosumers. The optimisation problem is formulated as follows.

\subsubsection{Upper Level Objective Function}
The DSO aims to maximise its revenue to ensure cost recovery of network investments. The objective is expressed as
\begin{equation}
    \underset{x^p \in \mathcal{X}^p, x^n \in \mathcal{X}^n}{\text{maximise}} \sum_{t \in \mathcal{T}} \sum_{i \in \mathcal{\mathcal{N}}} {p}^\text{g$+$}_{t,i} \pi^\text{g$+$}_{t}
    \label{eq: upper_obj}
\end{equation}
where $\mathcal{X}^p = \left\{p^{\text{g$+$}}_{t,i}, p^{\text{g$-$}}_{t,i}, p^{\text{g}}_{t,i}, p_{t,p}, p^{\text{c}}_{t,p}, p_{t,c}, \ell_{t,c}, p^{\text{$+$}}_{t,b}, p^{\text{$-$}}_{t,b}, e_{t,b} \right\}$ and $\mathcal{X}^n = \left\{\pi^{\text{g$+$}}_{t}, p_{t,g}, q_{t,g}, p_{t,lij}, q_{t,lij}, \ell_{t,l}, v_{t,i} \right\}$ represent the sets of prosumer and network decision variables, respectively. The set $\mathcal{T}$ denotes the time periods, and $\mathcal{N}$ represents the buses connected to prosumers. ${p}^\text{g$+$}_{t,i}$ is the power exported to the grid by prosumer at bus $i$ and time $t$, and $\pi^{\text{g$+$}}_{t}$ is the network tariff for power export at time $t$.

\subsubsection{Upper Level Constraints}

The amount that the DSO is permitted to charge is constrained by the annual repayment requirement for the cost recovery, formulated as 
\begin{IEEEeqnarray}{rCl}
     \sum_{t \in \mathcal{T}} \sum_{i \in \mathcal{N}} {p}^\text{g$+$}_{t,i} \pi^\text{g$+$}_{t} 
     \Delta t 
     &\leq& \Pi ,  
     \label{eq: upper_constr_revenue}
\end{IEEEeqnarray}
where the cap $\Pi$ can be readily adjusted to shorter horizons, such as daily limits or other resolutions, depending on the scope of analysis. \textcolor{black}{In this study, it is computed using a reference value of 1 AUD per customer per day, which is generally in line with typical fixed network tariffs in the Australian context. }$\Delta t$ is the step length.
In addition, constraints on active and reactive power balance are formulated as
\begin{IEEEeqnarray}{rCl}
    p_{t,lij} + p_{t,lji} &=& r_l \ell_{t,l}, \quad t \in \mathcal{T}, lij \in \mathcal{C},  \IEEEyesnumber\IEEEyessubnumber\\
    q_{t,lij} + q_{t,lji} &=& x_l \ell_{t,l}, \quad t \in \mathcal{T}, lij \in \mathcal{C}, \IEEEyessubnumber \\
    \sum_{gi \in \mathcal{C^{\text{gen}}}} p_{t,g} 
    &+& p^{\text{g}}_{t,i} 
    - \sum_{di \in \mathcal{C}^{\text{load}}}{\hat{p}^{\text{c}}_{t,d}}
    \nonumber\\
    &=& \sum_{lij \in \mathcal{C} \cup \mathcal{C}^{\text{rev}} } p_{t,lij}, \quad t \in \mathcal{T}, i \in \mathcal{I}, \IEEEyessubnumber \\
    \sum_{gi \in \mathcal{C}^{\text{gen}}} q_{t,g} 
    &-& \sum_{di \in \mathcal{C}^{\text{load}}}{\hat{q}^{\text{c}}_{t,d}} - \sum_{di \in \mathcal{C}^{\text{load}}}{\hat{q}^{\text{p}}_{t,d}} 
    \nonumber\\
    &=& \sum_{lij \in \mathcal{C} \cup \mathcal{C}^{\text{rev}}} q_{t,lij}, \quad t \in \mathcal{T}, i \in \mathcal{I}, \IEEEyessubnumber
\end{IEEEeqnarray}
where $\mathcal{C}$, $\mathcal{C}^{\text{rev}}$, $\mathcal{C}^{\text{gen}}$, and $\mathcal{C}^{\text{load}}$ represent line connectivity, lines with reversed bus order, generator connectivity, and load connectivity, respectively. 
$p_{t,lij}$ and $q_{t,lij}$ denote the active and reactive power flow leaving in line $l$ in the direction of node $i$ to $j$ at time $t$. $r_{l}$ and $x_{l}$ are the resistance and reactance of line $l$. $\ell_{t,l}$ represents the squared current magnitude flowing through line $l$ at time $t$. $p_{t,g}$ and $q_{t,g}$ are the active and reactive power generation of  generator $g$ for energy flexibility at time $t$. $\hat{p}^{\text{c}}_{t,d}$ stand for the consumer active demand forecast $d$ at time $t$, $\hat{q}^{\text{c}}_{t,d}$ and $\hat{q}^{\text{p}}_{t,d}$ denote the consumer and prosumer reactive demand forecast $d$ at time $t$. $p^{\text{g}}_{t,i}$ represents the net power exchange with the grid at bus $i$ and time $t$.
The voltage variable at bus $i$ and time $t$ is represented through symbols $v_{t,i}$, which represents the \emph{voltage magnitude squared}, i.e. $v_{t,i} = |V_{t,i}|^2$. 
We state Ohm's law between bus $i$ and $j$ in its lifted form,
\begin{IEEEeqnarray}{rCl}
    v_{t,j} &=& v_{t,i} - 2 \left( r_{l} p_{t,lij} + x_{l} q_{t,lij} \right)
    \nonumber\\
    &+& \left( r_{l}^2 + x_{l}^2 \right) \ell_{t,l}, \quad t \in \mathcal{T}, lij \in \mathcal{C}, 
\end{IEEEeqnarray}
The definition of power flow through the branch, i.e.  $p_{t,lij}^2 + q_{t,lij}^2 = \ell_{t,l} v_{t,i}$, is included through its conic relaxation, 
\begin{IEEEeqnarray}{rCl}
    \left\|   
    \begin{array}{c}
        2p_{t,lij} \\
        2q_{t,lij} \\
        \ell_{t,l} - v_{t,i}
    \end{array}
    \right\|_2 \leq \ell_{t,l} + v_{t,i}, \quad t \in \mathcal{T}, lij \in \mathcal{C}, \label{eq: upper_constr_soc}
\end{IEEEeqnarray}
which is exact for radial networks, provided that no upper bounds are imposed on loads \cite{Steven_2013_IEEEPowerSystems}.
Finally, the limits on the bus voltage magnitude are,
\begin{IEEEeqnarray}{rCl}
    (\underline{V}_{i})^2 \leq v_{t,i} \leq (\overline{V}_{i})^2, \quad t \in \mathcal{T}, i \in \mathcal{I},
\end{IEEEeqnarray}
where $\underline{V}_{i}$ and $\overline{V}_{i}$ denote the minimum and maximum voltage magnitude at bus $i$. 
The line current magnitude just has an upper bound,
\begin{IEEEeqnarray}{rCl}
    \ell_{t,l} \leq (\overline{I}_{l})^2, \quad t \in \mathcal{T}, l \in \mathcal{L},
\end{IEEEeqnarray}
where $\mathcal{L}$ is the set of distribution lines and $\overline{I}_{l}$ is the maximum current magnitude through line $l$. 
Line flow limits in terms of apparent power are,
\begin{IEEEeqnarray}{rCl}
    p_{t,lij}^2 + q_{t,lij}^2 &\leq& \overline{s}_{l}^2, \quad t \in \mathcal{T}, lij \in \mathcal{C} \cup \mathcal{C^{\text{rev}}}. 
\end{IEEEeqnarray}
with $\overline{s}_{l}$ represents the maximum apparent power flow through line $l$.

\subsection{Optimisation Problem for Prosumers}
As players in the lower level, the prosumers actively manage their behind-the-meter DERs according to the price signals. 

\subsubsection{Lower Level Objective Function}
The prosumers aim at minimising their electricity expenditure, and the objective function is,
\begin{equation}
\begin{aligned}
    \underset{x^p \in \mathcal{X}^p}{\text{minimise}} \quad 
    & \sum_{t \in \mathcal{T}} \sum_{i \in \mathcal{N}} \Big(
        p^{\text{g$-$}}_{t,i} \pi^{\text{WS}}_{t}
        - p^{\text{g$+$}}_{t,i} \pi^{\text{WS}}_{t}
        + p^{\text{g$+$}}_{t,i} \pi^{\text{g$+$}}_{t}
      \Big) 
      \Delta t  
      \\
    & \quad - \sum_{bi \in \mathcal{C}^{\text{BESS}}} 
        C^{e}_{b}\!\left(e_{T,b}\right),
\end{aligned}
\label{eq:lower_obj}
\end{equation}
where $\mathcal{C}^{\text{BESS}}$ denotes BESS connectivity. $p^{\text{g$-$}}_{t,i}$ represents the power imported from the grid by prosumer at bus $i$ and time $t$, $\pi^{\text{WS}}_t$ is the wholesale electricity price and time $t$. Prosumers are assumed to participate in the wholesale electricity market, purchasing and selling energy at wholesale rates, an operating mode inspired by Amber Electric \cite{amberelectric}. The term $p^{\text{g$-$}}_{t,i} \pi^{\text{WS}}_{t} \Delta t$ calculates the cost for importing energy from the grid, $p^{\text{g$+$}}_{t,i} \pi^{\text{WS}}_{t} \Delta t$ is the revenue for exporting energy to the grid, and $p^{\text{g$+$}}_{t,i} \pi^{\text{g$+$}}_{t} \Delta t$ is the amount of network charge paid according to the exported energy and the dynamic price. 
Function $C^e_b\left(e_{T,b}\right)$ is the battery energy end-value \cite{fredbattery}, formulated as 
\begin{IEEEeqnarray}{rCl}
    C^e_b\left(e_{T,b}\right) &=& \overline{M}_b \left(e_{T,b} - \frac{e^2_{T,b}}{2 \overline{e}_b} \right), b \in \mathcal{B},
\end{IEEEeqnarray}
where $\mathcal{B}$ is the set of BESS units. $\overline{M}_b$ denotes the marginal end-value at empty storage, $e_{T,b}$ is the final energy of battery $b$, and $\overline{e}_b$ is the maximum energy of battery $b$. This formulation effectively prevents batteries from being emptied at the end of the optimisation period and avoids the infeasibility issues that can arise when the terminal energy is fixed.

\subsubsection{Lower Level Constraints}

Prosumers' power exchange with the grid is expressed as
\begin{IEEEeqnarray}{rCl}
    p^{\text{g}}_{t,i} &=& \sum_{pi \in \mathcal{C^{\text{PV}}}}{p_{t,p}} - \sum_{ci \in \mathcal{C}^{\text{BESS}}} p_{t,c} 
    \nonumber\\
    &-& \sum_{di \in \mathcal{C}^{\text{load}}} \hat{p}^{\text{p}}_{t,d}, \quad t \in \mathcal{T}, i \in \mathcal{N}, \IEEEyesnumber\IEEEyessubnumber \\
    p^{\text{g}}_{t,i}  &=& p^{\text{g$+$}}_{t,i} - p^{\text{g$-$}}_{t,i}, \quad t \in T, i \in \mathcal{N}, \IEEEyessubnumber
\end{IEEEeqnarray}
where $\mathcal{C}^{\text{PV}}$ denotes PV system connectivity. $p_{t,p}$ is the PV generation after curtailment at panel $p$ and time $t$. $p_{t,c}$ denotes the active power flow from the grid into converter $c$ at time $t$. $p^{\text{g$+$}}_{t,i}$ and $p^{\text{g$-$}}_{t,i}$ denote the power exported to and imported from the grid by prosumer at bus $i$ and time $t$. The power exchange with the grid is limited by
\begin{IEEEeqnarray}{rCl}
    -\overline{p}^{\text{g$-$}}_{t,i} \leq p^{\text{g}}_{t,i} \leq \overline{p}^{\text{g$+$}}_{t,i}, \quad t \in T, i \in \mathcal{N} \IEEEyesnumber\IEEEyessubnumber \\
    p^{\text{g$+$}}_{t,i} \geq 0, p^{\text{g$-$}}_{t,i} \geq 0 , \quad t \in T, i \in \mathcal{N}. \IEEEyessubnumber
\end{IEEEeqnarray}
Power from the PV systems is expressed as
\begin{IEEEeqnarray}{rCl}
   p_{t,p} &=& \hat{p}_{t,p} - p^{\text{c}}_{t,p}, \quad t \in \mathcal{T}, p \in \mathcal{P}, \IEEEyesnumber\IEEEyessubnumber \\
    p_{t,p} &\geq& 0, p^{\text{c}}_{t,p} \geq 0, \quad t \in \mathcal{T}, p \in \mathcal{P}, \IEEEyessubnumber
\end{IEEEeqnarray}
where $\mathcal{P}$ denotes the set of PV systems. $\hat{p}_{t,p}$ \textcolor{black}{and $p^{\text{c}}_{t,p}$ are} the PV generation forecast \textcolor{black}{and PV curtailment} at panel $p$ and time $t$.
The constraints for battery converter power flow are formulated based on \cite{Fred_2020_cvt}, with only active power considered in this work for simplicity, 
\begin{IEEEeqnarray}{rCl}
    p_{t,c} ^2  &\leq& v_{t,j} \ell_{t,c}, \quad t \in \mathcal{T}, c \in \mathcal{B}, \IEEEyesnumber\IEEEyessubnumber \\
    p_{t,c} + p^{\text{$-$}}_{t,b} - p^{\text{$+$}}_{t,b} &=& r_c \ell_{t,c}, \quad t \in \mathcal{T}, c \in \mathcal{B}, \IEEEyessubnumber \\
    -\overline{p}_{c} &\leq& p_{t,c} \leq \overline{p}_{c}, \quad t \in \mathcal{T}, c \in \mathcal{B}, \IEEEyessubnumber
\end{IEEEeqnarray}
where $r_c$ is the converter resistance. $\ell_{t,c}$ is the squared current magnitude flowing into converter $c$ at time $t$. $p^{\text{$-$}}_{t,b}$ and $p^{\text{$+$}}_{t,b}$ are the discharging and charging power of battery $b$ at time $t$.
The batteries are modeled using the convex hull formulation \cite{batterychull} that captures the relationship between their energy and power, as well as the corresponding operational limits,

\begin{IEEEeqnarray}{rCl}
    e_{t,b} &=& e_{t-1,b} + \left(p^{\text{$+$}}_{t,b} \eta^{\text{$+$}}_{b} - p^{\text{$-$}}_{t,b} \frac{1}{\eta^{\text{$-$}}_b} \right) \Delta t, 
    \nonumber\\
    && \quad t \neq 1, t \in \mathcal{T}, b \in \mathcal{B}, \IEEEyesnumber\IEEEyessubnumber \\
    e_{t,b} &=& e^{\text{ini}}_b  + \left(p^{\text{$+$}}_{t,b} \eta^{\text{$+$}}_b - p^{\text{$-$}}_{t,b} \frac{1}{\eta^{\text{$-$}}_b} \right) \Delta t, 
    \nonumber\\
    &&\quad t=1, b \in \mathcal{B}, \IEEEyessubnumber \\
    0 &\leq& p^{\text{$+$}}_{t,b} \leq \overline{p}^{\text{$+$}}_b, \quad t \in \mathcal{T}, b \in \mathcal{B}, \IEEEyessubnumber \\
    0 &\leq& p^{\text{$-$}}_{t,b} \leq \overline{p}^{\text{$-$}}_b, \quad t \in \mathcal{T}, b \in \mathcal{B}, \IEEEyessubnumber \\
    \underline{e}_b &\leq& e_{t,b} \leq \overline{e}_b, \quad t \in \mathcal{T}, b \in \mathcal{B}, \IEEEyessubnumber \\
    \frac{p^{\text{$+$}}_{t,b}}{\overline{p}^{\text{$+$}}_b} &+& \frac{p^{\text{$-$}}_{t,b}}{\overline{p}^{\text{$-$}}_b} \leq 1, \quad t \in \mathcal{T}, b \in \mathcal{B}, \IEEEyessubnumber \\
    e_{t-1,b} &\leq& \overline{e}_b - \eta^{\text{$+$}}_b p^{\text{$+$}}_{t,b} \Delta t, \quad t \neq 1, t \in \mathcal{T}, b \in \mathcal{B}, \IEEEyessubnumber \\
    e^{\text{ini}}_b &\leq& \overline{e}_b - \eta^{\text{$+$}}_b p^{\text{$+$}}_{t,b} \Delta t, \quad t = 1, b \in \mathcal{B}, \IEEEyessubnumber \\
    e_{t-1,b} &\geq& \underline{e}_b + \frac{1}{\eta^{\text{$-$}}_b} p^{\text{$-$}}_{t,b} \Delta t, \quad t \in \mathcal{T}, b \in \mathcal{B}, \IEEEyessubnumber \\
    e^{\text{ini}}_b &\geq& \underline{e}_b + \frac{1}{\eta^{\text{$-$}}_b} p^{\text{$-$}}_{t,b} \Delta t, \quad t \in \mathcal{T}, b \in \mathcal{B}, \IEEEyessubnumber
\end{IEEEeqnarray}
where $e^{\text{ini}}_b$ is the initial energy of battery $b$. $\eta^{\text{$+$}}_b$ and $\eta^{\text{$-$}}_b$ denote the charging and discharging efficiency of battery $b$. $\overline{p}^{\text{$+$}}_b$ and $\overline{p}^{\text{$-$}}_b$ are the maximum charging and discharging power of battery $b$. $\underline{e}_b$ and $\overline{e}_b$ represent the minimum and maximum allowable energy of battery $b$.

\section{Case Study}
\label{casestudy}

\begin{figure}[tb]
\centering
\includegraphics[width=0.4\textwidth]{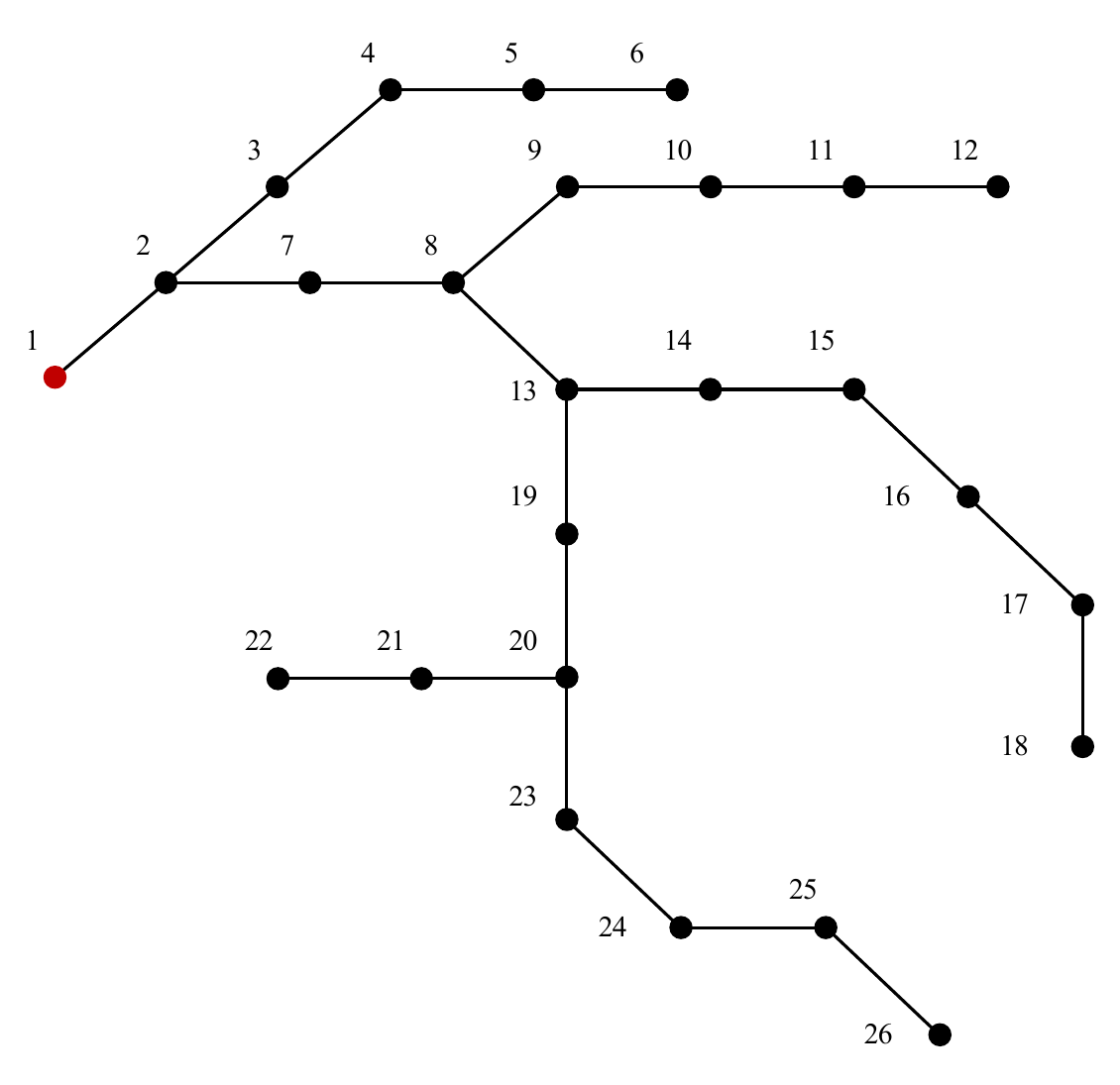}
\caption{Topology of the distribution network test system.}
\label{fig_testsystem}
\end{figure}

The performance of the proposed bilevel optimisation model is evaluated on a radial low-voltage distribution network, the topology of which is shown in Fig.~\ref{fig_testsystem}. A single-phase network is considered for simplicity, while the methodology can be extended to unbalanced three-phase systems; this will be addressed in future work. The test system comprises 25 prosumers equipped with PV systems and a subset with residential battery storage systems. Both PV and home batteries are operated at a unity power factor, while prosumer loads are assumed to draw power with an inductive power factor of 0.9. 
A DSO is also included as a player in the model, responsible for determining the dynamic network tariff, which is forwarded by the distribution company to consumers without modification. 
\textcolor{black}{
The lower-level problem \textcolor{black}{is formulated as} a convex quadratic objective and convex constraints. Using the \textsc{BiLevelJuMP.jl} \cite{garcia_2022_bjump} toolbox, we reformulate the bilevel problem by replacing the lower level with its necessary and sufficient KKT conditions, converting it to a single-level mathematical program with equilibrium constraints (MPEC). The bilinear terms in the upper level are handled through \textsc{Gurobi}'s branch-and-bound algorithm for nonconvex problems (enabled via the NonConvex=2 parameter).}
The implementation uses \textsc{Julia}~1.10.0 with \textsc{BiLevelJuMP}~0.6.2 for the reformulation and  \textsc{Gurobi}~12.0.2 as the underlying solver. The simulation setup and numerical results are presented and discussed in the following subsections.

\subsection{Simulation Setup}

\begin{figure}[tb]
    \centering
    \includegraphics[width=0.45 \textwidth]{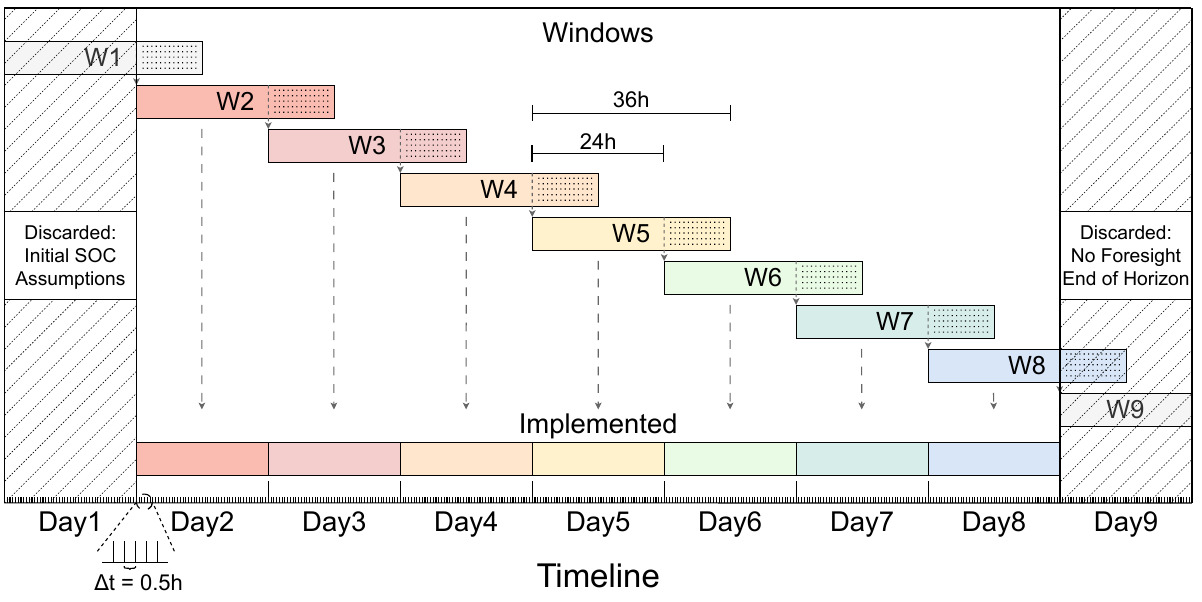}
    \caption{Schematic of the rolling horizon optimization framework.}
    \label{fig:rh}
\end{figure}

The simulation is carried out over a nine-day horizon using a rolling-horizon framework with a window length of 36 hours, a step size of 24 hours, and a time resolution $\Delta t$ of 30 minutes, 
\textcolor{black}{as illustrated in Fig.~\ref{fig:rh}.}
The first day is discarded, since the optimisation begins with assumed initial battery state of charge (SOC) values rather than the natural state that would result from prior operation. The final day is also excluded because the optimisation cannot extend beyond the planning horizon, and the last window may execute all slots without foresight, potentially leading to distorted SOC and price responses. Consequently, seven days of results are retained, ensuring realistic mid-horizon behaviour and avoiding edge artifacts.

Realistic Australian residential load profiles from the `Solar Analytics' dataset \cite{solaranalytics} have been adopted. We assume that all prosumers are equipped with a \SI{5}{\kilo\watt} PV system, while a portion are equipped with \SI{13.5}{\kilo\watt\hour} Tesla Powerwall 2 residential batteries initialised at \SI{50}{\percent} SOC. 
We constrain the SOC between the margins of \SI{20}{\percent} and \SI{80}{\percent} to preserve the headroom for the batteries to participate in other services.
Wholesale pricing signals, as well as the configuration and penetration levels of PV and battery systems, can be varied according to realistic scenarios without affecting the methodology. We assume that regulatory approval exists in the network region for grid-tied operation, and thus prosumers can discharge their batteries either to satisfy their local consumption or for export to the grid. We perform Monte Carlo analyses with random allocation of home batteries across the test system to validate the approach and to provide a general picture of the resulting price curves. Detailed input parameters are provided in Table \ref{table:inputparameters}.
\begin{table}[tb]
\caption{Input Parameters}
\label{table:inputparameters}
\renewcommand{\arraystretch}{1.2}
\centering
\begin{tabular}{
  >{\centering\arraybackslash}p{0.35\linewidth}
  >{\centering\arraybackslash}p{0.2\linewidth}
  >{\centering\arraybackslash}p{0.2\linewidth}}
\toprule
\textbf{Parameters} & \textbf{Unit} & \textbf{Value} \\
\midrule
$\underline{V}_{j}$, $\overline{V}_{j}$ & p.u. & 0.9, 1.1 \\
$\overline{I}_l$ & \si{\ampere} & 400 \\
$\overline{s}_l$ & \si{\kilo\volt\ampere} & 92\\
$\overline{p}^{\text{+}}_b$, $\overline{p}^{\text{$-$}}_b$ & \si{\kilo\watt} & 5.0 \\
$\eta^{\text{+}}_b$, $\eta^{\text{$-$}}_b$ & \si{\percent} & 95 \\
$\underline{e}_b$ & \si{\kilo\watt\hour} & 2.7 \\ 
$\overline{e}_b$ & \si{\kilo\watt\hour} & 10.8 \\
$r_c$ & \si{\ohm}  & 0.529 \\
$\hat{p}_{t,p}$ & \si{\kilo\watt} & [0-5] \\
$\pi^{\text{WS}}_t$ & \si{\$/\kilo\watt\hour} & 0.1 \\
$\Delta t$ & \si{\minute} & 30 \\
\bottomrule
\end{tabular}
\end{table}

\subsection{Numerical Results}
\begin{figure*}[!t]
    \centering
    \includegraphics[width=\textwidth]{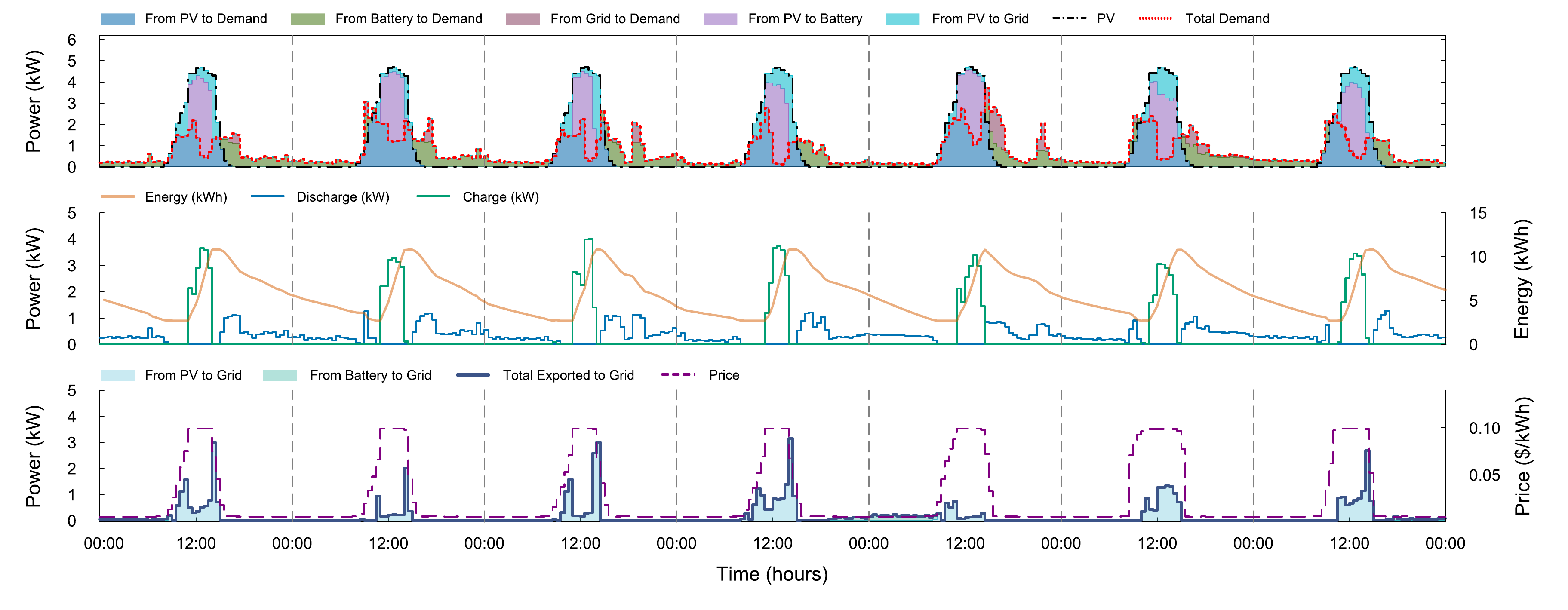}
    \caption{Simulation results for prosumer at Bus 4. From top to bottom are the illustration of demand composition, battery operation, power export with prices.}
    \label{fig:1_bus4_subplot}
\end{figure*}
\textcolor{black}{The numerical simulations were conducted on a workstation equipped with an Intel i7-12850HX CPU (2.10\,GHz) and 128 GB RAM. For the considered test system, typical solution times are approximately 15\,s per optimisation window. Incoporating uncertainty to capture limited or imperfect information about customer responses to price signals would increase problem size and may affect computation time, which is an important direction for future research.}

We analyse prosumer behaviour  including how local demand is met, the way batteries are operated, how exports occur in response to price signals, and the resulting expenditure. Network performance is also examined through voltage variation and line utilisation. 

\subsubsection{Supply and Demand Balancing}

The time-series operation of prosumers at Bus 4 over the one-week simulation shows how supply and demand are balanced using PV, BESS, and the grid. As illustrated in the first and second subplots in Fig.~\ref{fig:1_bus4_subplot}, during the morning and evening hours, when sunlight is absent and PV generation is zero, the demand is met primarily by battery discharge, the remainder supplied by grid imports. Around midday, PV generation becomes dominant, covering local demand and charging the battery. The battery is then discharged at other times to meet demand or export to the grid, while its energy level is kept within safe operating limits to avoid overcharging or deep discharge. 
Battery converter losses remain consistently below \SI{5}{\percent}, a typical value for residential-scale converters. Explicitly modelling converters and their associated losses is important, as it helps mitigate unrealistic charging and discharging oscillations and ensures smoother operational behaviour.
This coordinated interaction of PV, battery, and grid supply reflects the prosumers' objective of reducing electricity costs, as serving demand through PV generation and battery discharge is significantly less expensive than importing power from the grid at wholesale prices.

\subsubsection{Dynamic Tariff and Power Export}
In the third subplot, the blue area depicts the power exported to the grid, either from PV generation or battery discharge, while the purple curve represents the dynamic network export tariff calculated by the DSO and sent to the prosumers. Although the tariff varies daily due to differences in prosumer demand and operation, the overall trend is consistent: low at dawn, rising through the morning to approach but remain below wholesale rates, and then falling back to low levels at night. The midday increase corresponds to the period of significant PV generation across the network. From the prosumers’ perspective, higher tariffs reduce the net benefit of exporting to the grid, therefore incentivising storage of PV-generated energy in batteries for later use in satisfying demand, rather than immediate export. Conversely, when the network price is low, it encourages power export, and the prosumers choose to export their excess battery energy to the grid after satisfying their demand, as observed in the early hours of day one, during night of day two, in the morning of day three, and at night on day seven. Such exporting not only provides revenue for prosumers but also help ease network stress by reducing the burden of serving high loads.

\subsubsection{Other Prosumer Behaviour}
\begin{figure*}[!t]
    \centering
    \includegraphics[width=\textwidth]{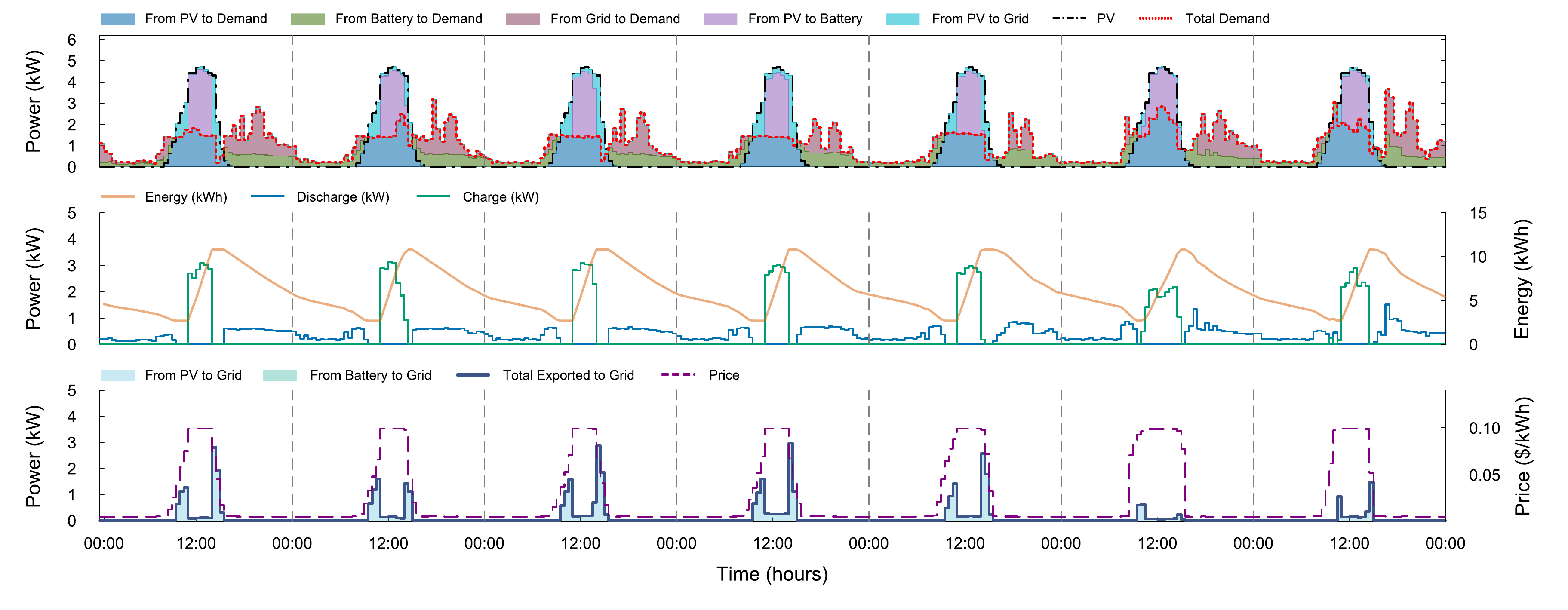}
    \caption{Simulation results for prosumer at Bus 12. From top to bottom are the illustration of demand composition, battery operation, power export with prices.}
    \label{fig:2_bus_12_subplot}
\end{figure*}
Fig.~\ref{fig:2_bus_12_subplot} illustrates the behaviour of prosumers at Bus 12, demonstrating the influence of local demand conditions on tariff responsiveness. Unlike Bus 4, where demand peaks around midday, Bus 12 exhibits a load profile concentrated at noon and in the afternoon. Here, the battery is discharged to satisfy local demand, with minimal residual energy available for grid export. Hence, the majority of exported power arises from surplus, non-curtailed PV generation at noon.

\subsubsection{Network Performance}
\begin{figure*}[!t]
\centering
\subfloat{\includegraphics[width=0.64\textwidth]{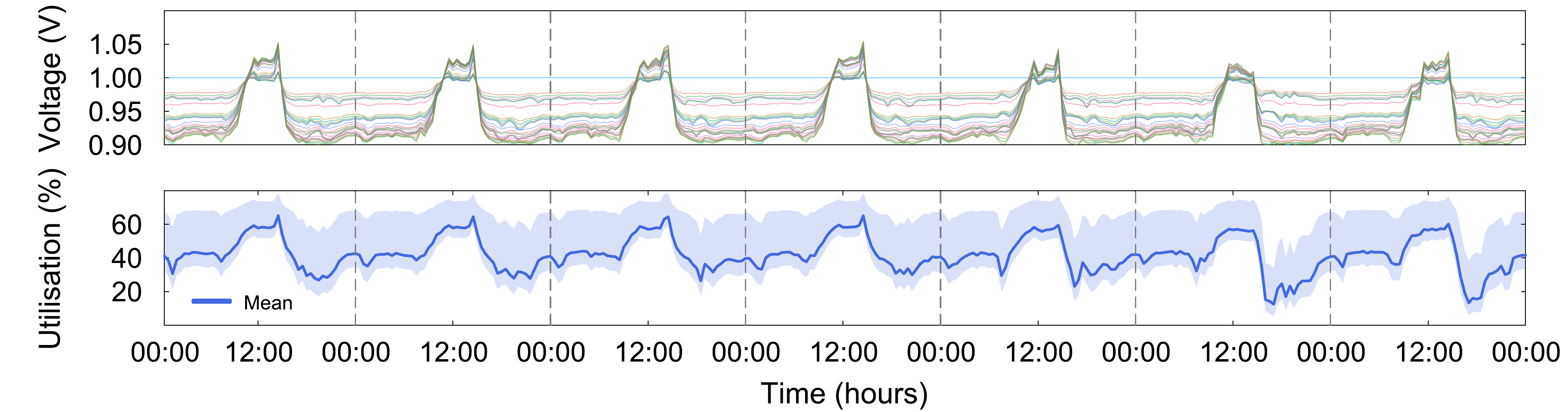}%
\label{1}}
\hspace{-0.01\textwidth}
\subfloat{\includegraphics[width=0.35\textwidth]{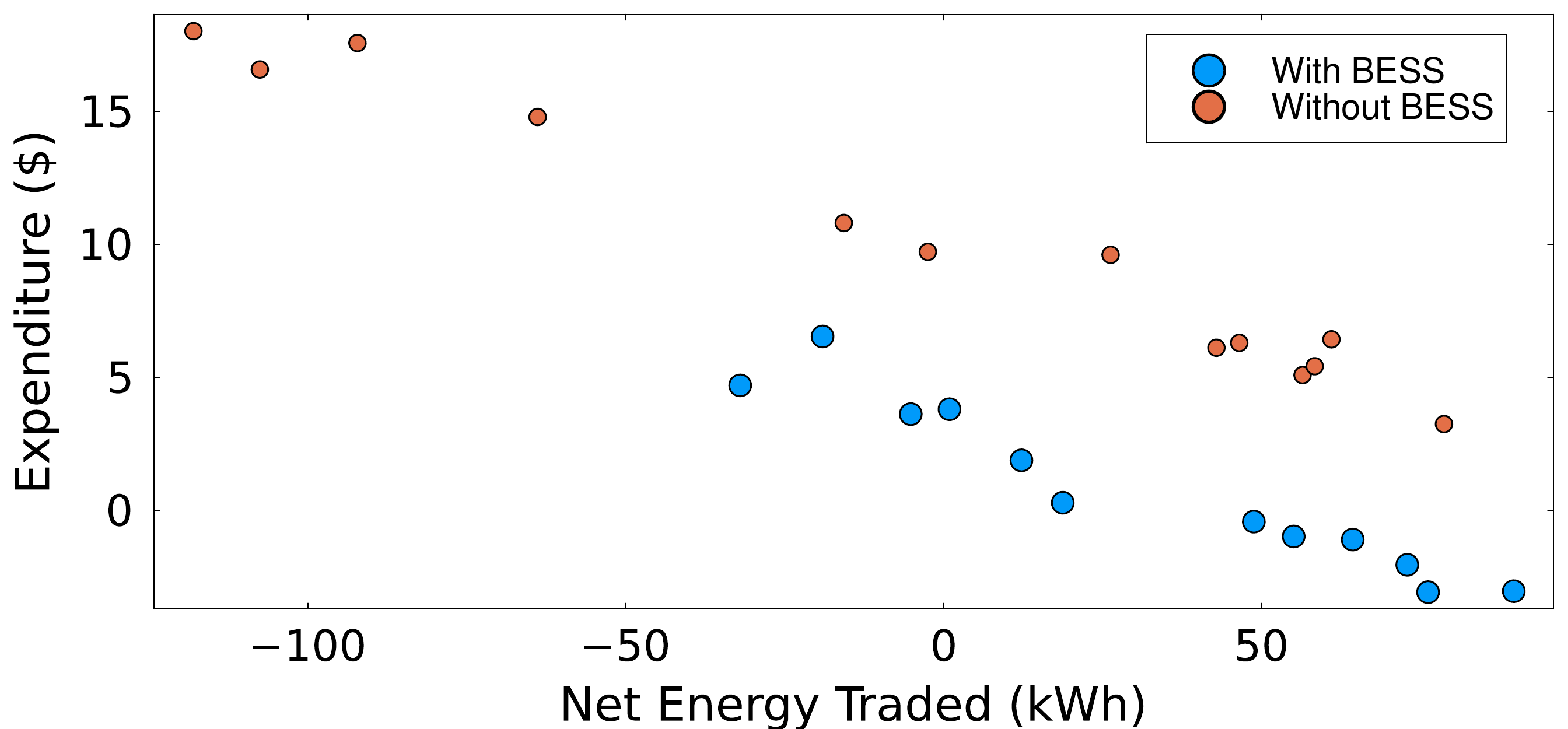}%
\label{2}}
\caption{The left panel presents voltage variation and line capacity utilisation over time. The right panel shows scatter plot for prosumer expenditure and net energy traded with/without BESS installation.}
\label{fig:3_5_line_expenditure}
\end{figure*}

The voltage at all prosumer buses and the line utilisation over the simulation horizon are depicted in the left panel of Fig.~\ref{fig:3_5_line_expenditure}. Bus voltages remain within the acceptable range of 0.9–1.1 p.u., rising at midday when exports peak and falling at other times due to local demand. Line utilisation stays below \SI{80}{\percent}, indicating no congestion under the simulated scenario. These results are consistent with the formulation of the optimisation model, which enforces constraints to ensure reliable network operation at the upper level.

\subsubsection{Prosumer Expenditure and Energy Traded}
Illustrated in the right panel of Fig.~\ref{fig:3_5_line_expenditure} is the prosumer expenditure versus their net energy traded, distinguishing households with and without BESS. Although energy usage patterns differ among houses, the results consistently indicate that prosumers with BESS obtain a clear advantage. Households without BESS show more positive net trade, reflecting their roles as net importers for most of the time, whereas those with BESS include both positive and negative values, demonstrating their ability to act as net exporters as well. This reflects the greater bidirectional flexibility of BESS-equipped houses, which can both absorb energy through charging and export strategically. By storing PV generation, they reduce dependence on grid imports and make fuller use of local production. As a result, for a given level of net energy traded, their expenditure is lower or even negative; conversely, for a given expenditure, those without BESS have to export more energy or import less to achieve comparable outcomes. By reducing reliance on imports and enabling controlled exports, BESS-equipped houses not only lower their own bills but also help flatten net demand on the network. This provides an economic incentive to encourage home battery installation as an additional benefit of the proposed model, offering advantages to both prosumers and the network.

\section{Discussion}
\label{discussion}

The proposed framework provides a methodology to design dynamic network prices that explicitly incorporates prosumer decision-making in hosting capacity management. The interaction can be interpreted as a Stackelberg game between the DSO and prosumers: the DSO first sets the export prices; prosumers then optimise their DER operation based on these prices and report back their responses; the DSO then updates prices accordingly. This iterative process continues until convergence, at which point the bilevel optimisation captures the equilibrium of the game.

Formulating the problem as a bilevel model avoids the common but unrealistic assumption that the DSO has full visibility of behind-the-meter activity or direct control of prosumer assets. This separation reflects regulatory and privacy limitations while providing economic incentives that safeguard network reliability and align the interests of both DSOs and prosumers.
\textcolor{black}{
In addition, since loads are upper-bounded in our paper, the SOC\textcolor{black}{P} power flow relaxation can occasionally be inexact. This inexactness appears as \textit{fictitious demand}, produced by increased network losses and simultaneous battery charging and discharging. Leaving PV curtailment substantially reduces both the frequency and severity of these artifacts, although it does not make relaxation perfectly tight. In other words, if PV curtailment were not available, more instances of relaxation inexactness would occur. Overall, the residual impact on our results is negligible.
}

\section{Conclusion}
\label{conclusion}

This paper has presented a bilevel optimisation framework for designing dynamic network prices, offering a novel approach to managing the LV network hosting capacity under high DER penetration. The proposed approach separates the roles and objectives of DSO and prosumers, ensuring equilibrium consistency and maintaining customer prerogative. By linking network prices to local operating conditions, the framework supports coordinated DER operation and efficient utilisation of network capacity. The results demonstrate that dynamic network prices provide effective economic incentives for prosumer participation, reducing export peaks, and facilitating reliable network operation. Fairness is also preserved, as all prosumers receive identical price signals regardless of their location within the network. The formulation provides a practical basis for implementing price-based network hosting capacity management strategies in LV systems, informing the design of dynamic network tariffs, and supporting the transition to economically efficient and customer-centric operation.

\bibliographystyle{IEEEtran}
\bibliography{references}

@report{2024_Roadmap,
    author={{Department of Climate Change, Energy, the Environment and Water}},
    title={{National Consumer Energy Resources  Roadmap}},
    year={2024}
}

@article{Ismael_2019_RenewableEnergy,
author = {Ismael, Sherif and Abdel Aleem, Shady and Abdelaziz, Almoataz and Zobaa, Ahmed},
year = {2019},
month = {01},
pages = {1002-1020},
title = {State-of-the-art of Hosting Capacity in Modern Power Systems with Distributed Generation},
volume = {130},
journal = {Renew. Energy},
doi = {10.1016/j.renene.2018.07.008}
}

@article{Koirala_2022_RSER,
title = {Hosting capacity of photovoltaic systems in low voltage distribution systems: A benchmark of deterministic and stochastic approaches},
journal = {Renew. Sust. Energy Rev.},
volume = {155},
pages = {111899},
year = {2022},
issn = {1364-0321},
doi = {https://doi.org/10.1016/j.rser.2021.111899},
author = {Arpan Koirala and Tom {Van Acker} and Reinhilde D’hulst and Dirk {Van Hertem}},
}

@report{ARENA_2022_DOE,
    author={{Australian Renewable Energy Agency (ARENA)}},
    title={{Review of Dynamic Operating Envelope Adoption by DNSPs}},
    url={https://arena.gov.au/assets/2022/07/review-of-dynamic-operating-envelopes-from-dnsps.pdf},
    year={2022}
}

@report{Edith_2022_ENEA,
    author={{Enea Consulting}},
    title={{Project Edith Overview Report}},
    url={https://www.ausgrid.com.au/-/media/Documents/Reports-and-Research/Project-Edith/Project-Edith-2022.pdf},
    year={2022}
}

@report{ACCC_2024_Tariffs,
    author={{Australian Competition and Consumer Commission}},
    title={{Inquiry into the National Electricity Market}},
    year={2024}
}

@article{Iacopo_2021_Omega,
title = {Electricity prices and tariffs to keep everyone happy: A framework for fixed and nodal prices coexistence in distribution grids with optimal tariffs for investment cost recovery},
journal = {Omega},
volume = {103},
pages = {102450},
year = {2021},
issn = {0305-0483},
doi = {https://doi.org/10.1016/j.omega.2021.102450},
author = {Iacopo Savelli and Thomas Morstyn},
}

@ARTICLE{Hayat_2019_IEEEIndustrialInformatics,
  author={Hayat, Muhammad Adnan and Shahnia, Farhad and Shafiullah, GM},
  journal={IEEE Trans. Indust. Informatics}, 
  title={Replacing Flat Rate Feed-In Tariffs for Rooftop Photovoltaic Systems With a Dynamic One to Consider Technical, Environmental, Social, and Geographical Factors}, 
  year={2019},
  volume={15},
  number={7},
  pages={3831-3844},
  doi={10.1109/TII.2018.2887281}}

@ARTICLE{Steven_2013_IEEEPowerSystems,
  author={Farivar, Masoud and Low, Steven H.},
  journal={IEEE Trans. Power Syst.}, 
  title={Branch Flow Model: Relaxations and Convexification - {Part I}}, 
  year={2013},
  volume={28},
  number={3},
  pages={2554-2564},
  doi={10.1109/TPWRS.2013.2255317}}

@misc{amberelectric,
  url = {https://www.amber.com.au/},
  author={{Amber Electric}}
}

@misc{solaranalytics,
  url = {https://www.solaranalytics.com.au/},
  author={{Solar Analytics}}
}

@article{fredbattery,
title = {Non-linear charge-based battery storage optimization model with bi-variate cubic spline constraints},
journal = {J. Energy Storage},
volume = {32},
pages = {101979},
year = {2020},
issn = {2352-152X},
doi = {https://doi.org/10.1016/j.est.2020.101979},
author = {Per Aaslid and Frederik Geth and Magnus Korpås and Michael M Belsnes and Olav B Fosso}}

@ARTICLE{batterychull,
  author={Pozo, David},
  journal={IEEE Trans. Power Syst.}, 
  title={Convex Hull Formulations for Linear Modeling of Energy Storage Systems}, 
  year={2023},
  volume={38},
  number={6},
  pages={5934-5936},
  doi={10.1109/TPWRS.2023.3304131}}

@inproceedings{Fred_2020_cvt,
author = {Geth, Frederik and Coffrin, Carleton and Fobes, David},
title = {A Flexible Storage Model for Power Network Optimization},
year = {2020},
isbn = {9781450380096},
publisher = {Association for Computing Machinery},
address = {New York, NY, USA},
doi = {10.1145/3396851.3402121},
booktitle = {Proceedings of the Eleventh ACM International Conference on Future Energy Systems},
location = {Virtual Event, Australia},
series = {e-Energy '20}
}

@article{Afshin_2025_SEGAN,
title = {Social welfare maximization in unbalanced distribution networks under dynamic pricing and power exchange limits},
journal = {Sust. Energy Grids Networks},
volume = {44},
pages = {101923},
year = {2025},
issn = {2352-4677},
doi = {https://doi.org/10.1016/j.segan.2025.101923},
author = {Afshin Najafi-Ghalelou and Mohsen Khorasany and M.Imran Azim and Reza Razzaghi},
}

@online{Acacia_bill,
  author = {{Acacia Energy}},
  title = {Understanding your electricity bill},
  url = {https://www.acaciaenergy.com.au/blog/understanding-your-electricity-bill/},
}

@misc{Powercor_202526,
  author = {{Powercor Australia}},
  title = {Powercor Tariff Summary 2025-26}, 
  year = {2025}
}

@misc{garcia_2022_bjump,
      title={{BilevelJuMP.jl}: Modeling and Solving Bilevel Optimization in {Julia}}, 
      author={Joaquim Dias Garcia and Guilherme Bodin and Alexandre Street},
      year={2022},
      eprint={2205.02307},
      archivePrefix={arXiv},
      primaryClass={math.OC},
      url={https://arxiv.org/abs/2205.02307}, 
}

@ARTICLE{Papavasiliou_2018_IEEESmartGrid,
  author={Papavasiliou, Anthony},
  journal={IEEE Trans. Smart Grid}, 
  title={Analysis of Distribution Locational Marginal Prices}, 
  year={2018},
  volume={9},
  number={5},
  pages={4872-4882},
  doi={10.1109/TSG.2017.2673860}}

@ARTICLE{Bai_2018_IEEEPowerSystems,
  author={Bai, Linquan and Wang, Jianhui and Wang, Chengshan and Chen, Chen and Li, Fangxing},
  journal={IEEE Trans. Power Syst.}, 
  title={{Distribution Locational Marginal Pricing (DLMP) for Congestion Management and Voltage Support}}, 
  year={2018},
  volume={33},
  number={4},
  pages={4061-4073},
  doi={10.1109/TPWRS.2017.2767632}}

@ARTICLE{Alahmed_2025_IEEEContrlNS,
  author={Alahmed, Ahmed S. and Cavraro, Guido and Bernstein, Andrey and Tong, Lang},
  journal={IEEE Trans. Control Netw. Syst.}, 
  title={A Decentralized Market Mechanism for Energy Communities Under Operating Envelopes}, 
  year={2025},
  volume={12},
  number={1},
  pages={313-324},
  doi={10.1109/TCNS.2024.3466651}}

\end{document}